\documentclass[10pt, doublecolumn]{IEEEtran}
\usepackage{graphicx}
\usepackage{caption}
\usepackage[ruled]{algorithm2e}
\ifCLASSOPTIONcompsoc
  \usepackage[caption=false,font=normalsize,labelfont=sf,textfont=sf]{subfig}
\else
  \usepackage[caption=false,font=footnotesize]{subfig}
\fi
\usepackage{indentfirst}

\usepackage{amsmath}
\allowdisplaybreaks[4]
\usepackage{amssymb}
\usepackage{times}
\usepackage{mathtools}
\usepackage{psfrag}
\usepackage{cite}
\usepackage{lastpage}
\usepackage{fancyhdr}
\usepackage{color}
 \usepackage{amsthm}
\usepackage{bigints}
\def\Pr{\text{Pr}}

\newtheorem{Lemma}{Lemma}
\theoremstyle{remark}
\newtheorem{Remark}{$\quad$Remark}

\begin{document}
\title{A Fine Grained Stochastic Geometry Based Analysis on LEO Satellite Communication Systems}
\author{Yanshi Sun, \IEEEmembership{Member, IEEE}, Zhiguo Ding, \IEEEmembership{Fellow, IEEE}
\thanks{
The work of Y. Sun  is supported by Hefei University of Technology's construction funds for the introduction of talents with funding number 13020-03712022011.

Y. Sun is with the School of Computer Science and Information
Engineering, Hefei University of Technology, Hefei, 230009, China. (email: sys@hfut.edu.cn).
Z. Ding is with Department of Electrical Engineering and Computer Science, Khalifa University, Abu Dhabi, UAE, and Department of Electrical and Electronic Engineering, University of Manchester, Manchester, UK. (email:zhiguo.ding@manchester.ac.uk).
}
\vspace{-2.5em}}
\maketitle
\begin{abstract}
Recently, stochastic geometry has been applied to provide tractable performance analysis for low earth orbit (LEO) satellite networks. However, existing works mainly focus on analyzing the ``coverage probability'', which provides limited information. To provide more insights, this paper provides a more fine grained analysis on LEO satellite networks modeled by a homogeneous Poisson point process (HPPP). Specifically, the distribution and moments of the conditional coverage probability given the point process are studied. The developed analytical results can provide characterizations on LEO satellite networks, which are not available in existing literature, such as ``user fairness'' and ``what fraction of users can achieve a given transmission reliability ''. Simulation results are provided to verify the developed analysis. Numerical results show that, in a dense satellite network, {\color{black}it is} beneficial to deploy satellites at low altitude, for the sake of both
coverage probability and user fairness.
\end{abstract}
\section{Introduction}
{\color{black}Recently, deploying low earth orbit (LEO) satellite constellations to provide
ubiquitous global connectivity is becoming an important enabling technique for
6G wireless communications \cite{giordani2020non}.
Because LEO satellite constellations require a dense deployment of satellites, performance evaluation using computer simulations can be time-consuming but yeild limited insight.
To address this challenge, emerging research is exploring the use of tools from stochastic geometry \cite{haenggi2012stochastic} to evaluate the performance of LEO satellite networks theoretically and provide a better understanding of their properties \cite{al2021analytic, al2021modeling, lee2022coverage, okati2020downlink, talgat2020stochastic}.
It is worth noting that existing research in this area typically focuses on metrics that reveal the average performance of the entire network, such as ``the coverage probability''.
However, many important properties of satellite communication networks remain unclear. For example, we may ask: i) how does a satellite network perform in terms of user fairness? ii) what fraction of users in the network can achieve  certain link reliability?}

To answer the above questions, this letter aims to provide a more fine grained stochastic geometric performance analysis for downlink LEO satellite networks.
Specifically, LEO satellites, which provide service to ground users, are modeled as a homogeneous Poisson point process (HPPP) denoted by $\Phi$.
The conditional coverage probability given $\Phi$ with respect to a signal-to-interference-ratio (SIR) threshold $\theta$, denoted by $P_s(\theta)$, is first evaluated. Note that $P_s(\theta)$ itself is a random variable driven by $\Phi$. Different from existing work which only studies the mean value of $P_s(\theta)$, this paper investigates the moments and distribution
of $P_s(\theta)$.  The developed analytical results can provide characterization on
``user fairness'' and ``what fraction of users can achieve a given transmission reliability ''.
Numerical results show that, for a dense satellite, it is better to deploy satellites at low altitude, by jointly considering users' average performance and user fairness. While for a sparser network, a high altitude is more favorable for deploying satellites.
\section{System Model}
\begin{figure}[!t]
\vspace{-0em}
\setlength{\abovecaptionskip}{0em}   %
\setlength{\belowcaptionskip}{-2em}   %
  \centering
  \includegraphics[width=2.5in]{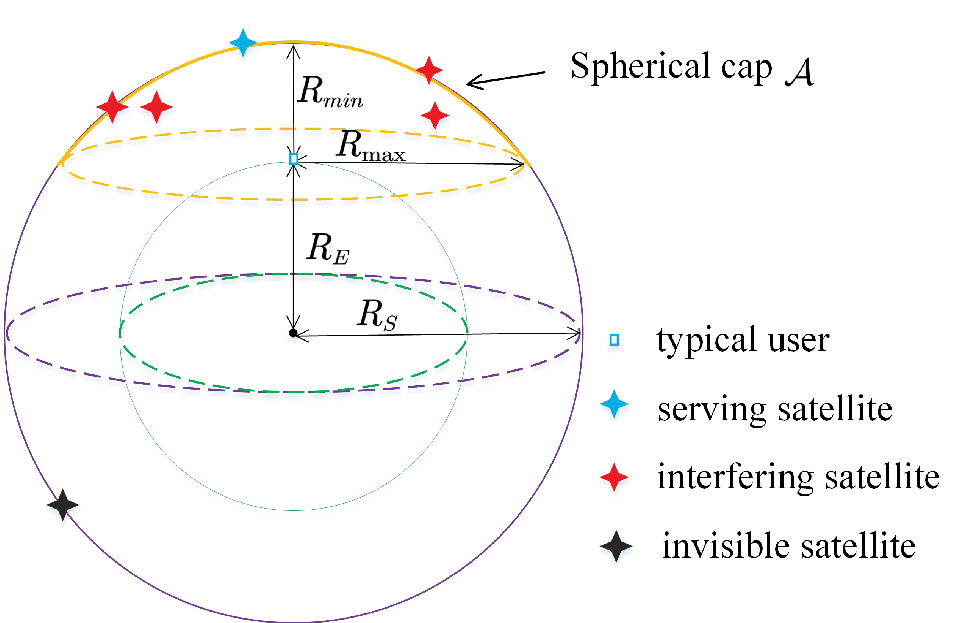}\\
  \caption{Illustration of the system model.}
  \label{system_model}
\end{figure}
Consider a downlink satellite communication scenario as shown in Fig. \ref{system_model}.
The earth is modeled as a sphere with radius $R_E$.
The LEO satellites are flying around the earth with circular orbit at the same altitude $R_{min}$ above the mean sea level.
For a given short time slot, the locations of the satellite can be assumed to be fixed, and the focus of this paper is to evaluate the performance achieved by the ground users served by satellites in such a short time slot. For a given short time slot, it is reasonable to assume that the locations of the LEO satellites form a homogeneous Poisson Point Process (HPPP) on the surface of the sphere (denoted by $\mathbb{S}^2_S$) centered at the geocentric with radius $R_S=R_E+R_{min}$, denoted by $\Phi=\{x_i\}$, where $x_i$ is the coordinate of the $i$-th satellite.
The intensity of $\Phi$ is denoted by $\lambda$, which indicates the density of the satellites.

Ground users can also be modeled as a HPPP with intensity $\lambda_u$ on the surface of the earth. Due to the stationarity of HPPP, it is sufficient to consider a typical ground user $U_0$ located at $(0,0,R_E)$. It is assumed that each ground user is served by its nearest satellite\footnote{The nearest satellite has to be within the horizon of the user, otherwise {\color{black}there is} no satellite serving the user.}, and all other satellites which are within the horizon of the user play the role of interference sources. For the ease of exposition, define a
spherical cap $\mathcal{A}$ which is
the portion of sphere surface $\mathbb{S}^2_S$ cut off by a tangent plane to the earth surface
at $(0, 0,R_E)$. According to the previous discussion, only the satellites located on $\mathcal{A}$
affect the performance of $U_0$.

In the rest of this paper, the satellites are ordered according to their distances to $U_0$, i.e., $x_1$ is nearest satellite's coordinate.
Given that {\color{black}there is} at least one satellite on $\mathcal{A}$, the SIR at $U_0$ is given by:
\begin{align}
 \text{SIR}&=\frac{|h_1|^2}{\sum_{x_i\in \Phi\backslash\{x_1\}\cap \mathcal{A}}|h_i|^2}
            \overset{\Delta}{=}\frac{|h_1|^2}{I},
\end{align}
where $h_i=g_ir_i^{-\alpha/2}$ is the channel between $U_0$ and the $i$-th satellite \cite{an2016secure,lin2022slnr,lin2022refracting}, $g_i$ is the small scale fading, which is modeled as Nakagami fading with parameter $M$. Consequently, $|g_i|^2$ is a normalized Gamma
random variable with parameter $M$,  whose
cumulative density function (CDF) is given by:
\begin{align}
 F_{|g_i|^2}(x)=\gamma(M,Mx)/\Gamma(M)
               =1-\sum_{i=0}^M\frac{(Mx)^i}{i!}e^{-Mx},
\end{align}
where $\Gamma(M)=(M-1)!$ and $\gamma(s,x)$ is the lower incomplete gamma function, given by
$
 \gamma(s,x)=\int_{0}^xt^{s-1}e^{-t}\,dt.
$
$r_i$ is the distance between $U_0$ and the $i$-th satellite. Note that for any satellite within area $\mathcal{A}$, $r_i$ is upper bounded by $R_{max}=\sqrt{R_S^2-R_E^2}$.
And $\alpha$ is the large scale path loss exponent.
{\color{black}It is assumed that each ground user knows perfect channel state information (CSI) to its serving satellite.}

And $I=\sum_{x_i\in \Phi\backslash\{x_1\}\cap \mathcal{A}}|h_i|^2$ is the sum of the interferences.

In the following, we would like to characterize the distribution of SIR. Note that there are two kinds of randomness which impact SIR, one is the small scale fading, and the other is $\Phi$. Given $\Phi$, the conditional coverage probability that the SIR is beyond a threshold $\theta$ is given by:
\begin{align}
 P_s(\theta)&\overset{\Delta}{=}\text{Pr}\left(\text{SIR}>\theta, \Phi(\mathcal{A})>0|\Phi \right)\\\notag
            &=\text{Pr}\left(\text{SIR}>\theta|\Phi,\Phi(\mathcal{A})>0 \right)\mathbf{1}(\Phi(\mathcal{A})>0|\Phi),
\end{align}
where $\mathbf{1}(\cdot)$ is the indicator function.
Note that $P_s(\theta)$ is a random variable whose distribution depends on the distribution of $\Phi$.
In the existing literature, only the mean value of $P_s(\theta)$, i.e., the coverage probability of the  typical link, is studied. The ``coverage probability''  can only reflect the average performance of
all users, which is just a rough characterization of the system performance.
This paper will give a more fine grained characterization of $P_s(\theta)$,
to provide more statistical insights, in the next section.

\section{Performance Analysis}
In this section, the moments of $P_s(\theta)$ are first evaluated, then the meta distribution, i.e., {\color{black}the complementary cumulative distribution function (CCDF)} of $P_s(\theta)$,  is approximated via beta approximation by using the first and second order moments of $P_s(\theta)$.
\begin{Lemma}
Given $\Phi$ and $\Phi(\mathcal{A})>0$, the conditional coverage probability that
the SIR is larger than the threshold $\theta$ can be approximated as:
\begin{align}\label{con_cov_Pr}
  &\quad\Pr\left(\text{SIR}>\theta|\Phi,\Phi(\mathcal{A})>0 \right)\\\notag
   &\approx \sum_{m=1}^{M}C_{M}^m(-1)^{m+1}\prod\limits_{x_{i}\in\Phi\backslash\{x_1\}\cap{\mathcal{A}}}\frac{1}{
   \left(1+\frac{m\eta\theta r_1^{\alpha}}{Mr_i^{\alpha}}\right)^M}.
\end{align}
\end{Lemma}
\begin{IEEEproof}
The conditional coverage probability can be calculated as follows:
\begin{align}
   &\quad\Pr\left(\text{SIR}>\theta|\Phi,\Phi(\mathcal{A})>0 \right)\\\notag
   &=\Pr\left(|g_1|^2>\theta r_1^{\alpha}I|\Phi,\Phi(\mathcal{A})>0\right)\\\notag\
   &=1-\Pr\left(|g_1|^2<\theta r_1^{\alpha}I|\Phi,\Phi(\mathcal{A})>0\right)\\\notag
   &\overset{(a)}{\approx} 1-\mathbb{E}_{g_i}\left\{\left(1-e^{-\eta\theta r_1^\alpha I}\right)^M |\Phi,\Phi(\mathcal{A})>0\right\}\\\notag
   &\overset{(b)}{=}\mathbb{E}_{g_i}\left\{\sum_{m=1}^{M}C_{M}^m(-1)^{m+1}e^{-m\eta\theta r_1^{\alpha}I} |\Phi,\Phi(\mathcal{A})>0\right\}\\\notag
   &=\mathbb{E}_{g_i}\left\{\sum_{m=1}^{M}C_{M}^m(-1)^{m+1}\prod\limits_{x_{i}\in\Phi\backslash\{x_1\}\cap{\mathcal{A}}}\right.\\\notag
   &\quad\quad\quad\quad\quad\quad\quad\left.\left.e^{-\frac{-m\eta\theta r_1^{\alpha}|g_i|^2}{r_i^{\alpha}}}\right|\Phi,\Phi(\mathcal{A})>0\right\}\\\notag
   &\overset{(c)}{=}\sum_{m=1}^{M}C_{M}^m(-1)^{m+1}\prod\limits_{x_{i}\in\Phi\backslash\{x_1\}\cap{\mathcal{A}}}\frac{1}{
   \left(1+\frac{m\eta\theta r_1^{\alpha}}{Mr_i^{\alpha}}\right)^M},
\end{align}
where step (a) follows from the fact that the CDF of the normalized gamma random variable $|g_1|^2$
can be tightly lower bounded by \cite{alzer1997some}:
\begin{align}\label{inEQ}
F_{|g_1|^2}(x)\geq \left(1-e^{-\eta x}\right)^M,
\end{align}
where $\eta=M(M!)^{-\frac{1}{M}}$, and the lower bound is used to approximate the probability; step (b) follows from applying binomial expansion; and step (c) follows from taking average with respect to the small scale fadings of the interfering links, which are independently and identically distributed (i.i.d) normalized gamma random variables with parameter $M$.
\end{IEEEproof}

\begin{Remark}
Note that when $M=1$, the small scale fading degrades to Rayleigh fading, and  equality holds for (\ref{inEQ}), resulting in no approximation error in (\ref{con_cov_Pr}).
\end{Remark}

The $b$-th moment of $P_s(\theta)$ is defined as:
\begin{align}
 M_b(\theta)=\mathbb{E}_{\Phi}\{P_s^b(\theta)\}
\end{align}

In the next, we would like to provide expressions for $M_{b}(\theta)$. To this end, it is necessary to obtain the probability density function (PDF) of $r_1$ given that $\Phi(\mathcal{A})>0$, denoted by $f_{r_1|\Phi(\mathcal{A})>0}(r)$. According to \cite{lee2022coverage}, the conditional PDF can be expressed as:
\begin{align}
 f_{r_1|\Phi(\mathcal{A})>0}(r)=\upsilon(\lambda,R_S)re^{-\lambda\pi\frac{R_S}{R_E}r^2}, R_{min}\leq r \leq R_{max},
\end{align}
where $\upsilon(\lambda,R_S)=2\pi\lambda\frac{R_S}{R_E}\frac{e^{\lambda\pi\frac{R_S}{R_E}(R_S^2-R_E^2)}}{
e^{2\pi\lambda R_S(R_S-R_E)}-1}$.

\begin{Lemma}
For a positive integer $b$, the $b$-th moment of $P_s(\theta)$ can be expressed as follows:
\begin{align}
  M_b(\theta)\approx G(\theta)\left(1-\exp\left(-2\pi\lambda R_{min}R_S \right)\right),
\end{align}
where
\begin{align}
 &G(\theta)\approx \sum_{b_1,\cdots,b_M}{b \choose  b_1,\cdots,b_M} \prod_{m=1}^{M}\left(C_{M}^{m}(-1)^{m+1}\right)^{b_m}\\\notag
  &\frac{(R_{max}\!\!-\!\!R_{min})\pi}{2N}\!\sum_{k=1}^{K}\!\sqrt{1\!\!-\!\psi_k^2}
  \exp\left(-Q(d_k,\theta)\right)f_{r_1|\Phi(\mathcal{A})>0}(d_k),
\end{align}
$K$ is the Gaussian-Chebyshev approximation \cite{sys2019PCP} parameter, $\psi_k=\cos{\frac{(2k-1)\pi}{2K}}$,
$d_k=\frac{R_{max}-R_{min}}{2}\psi_k+\frac{R_{max}+R_{min}}{2}$,
and
\begin{align}
 Q(r_1,\theta)\approx &\frac{\pi^2\lambda R_S(R_{max}-r_1)}{NR_E}\sum_{n=1}^N\sqrt{1-\phi_n^2}
 c_n\\\notag
 &\times\left(1-\prod_{m=1}^{M}{\left(1+\frac{m\eta\theta r_1^{\alpha}}{Mc_n^{\alpha}}\right)^{-Mb_m}}\right)
\end{align}
where $N$ is the Gaussian-Chebyshev approximation parameter, $\phi_n=\cos{\frac{(2n-1)\pi}{2N}}$,
$c_n=\frac{R_{max}-r_1}{2}\phi_n+\frac{R_{max}+r_1}{2}$.
\end{Lemma}

\begin{IEEEproof}
According to the definition of $M_b(\theta)$, it can be expressed as follows:
\begin{align}
  M_b(\theta)=\mathbb{E}_{\Phi|\Phi(\mathcal{A})>0}\{P_s^b(\theta)\}\Pr(\Phi(\mathcal{A})>0).
\end{align}
Note that $\Pr(\Phi(\mathcal{A})>0)$ is the probability that {\color{black}there is} at least one satellite on the spherical cap $\mathcal{A}$, which can be easily obtained according to the properties of HPPP, as follows \cite{lee2022coverage}:
\begin{align}
 \Pr(\Phi(\mathcal{A})>0)=1-\exp\left(-2\pi\lambda R_{min}R_S \right).
\end{align}
The remaining  task is to evaluate $\mathbb{E}_{\Phi|\Phi(\mathcal{A})>0}\{P_s^b(\theta)\}\overset{\Delta}{=}G(\theta)$. By applying Lemma $1$, $G(\theta)$ can be expressed as:
\begin{align}
 G(\theta)=&\mathbb{E}_{\Phi|\Phi(\mathcal{A})>0}\left\{\left(\sum_{m=1}^{M}C_{M}^m(-1)^{m+1}\right.\right.\\\notag
 &\left.\left.\prod\limits_{x_{i}\in\Phi\backslash\{x_1\}\cap{\mathcal{A}}}\frac{1}{
   \left(1+\frac{m\eta\theta r_1^{\alpha}}{Mr_i^{\alpha}}\right)^M}\right)^b\right\},
\end{align}
By applying polynomial expansion, $G(\theta)$ can be further expressed as:
\begin{align}
 &G(\theta)=\mathbb{E}_{\Phi|\Phi(\mathcal{A})>0}\left\{
   \sum_{b_1,\cdots,b_M}{b \choose b_1,\cdots,b_M} \prod_{m=1}^M \right.\\\notag
   &\left.\quad\quad\left(C_{M}^m(-1)^{m+1}\right)^{b_m} \!\!\!\!\!\!\prod\limits_{x_{i}\in\Phi\backslash\{x_1\}\cap{\mathcal{A}}}
   \prod_{m=1}^M\frac{1}{\left(1+\frac{m\eta\theta r_1^{\alpha}}{M r_i^{\alpha}}\right)^{Mb_m}}
 \right\}.
\end{align}
Then, by applying the probability generating functional (PGFL) of HPPP \cite{haenggi2012stochastic}, it is obtained that
\begin{align}
 &G(\theta)=\mathbb{E}_{r_1|\Phi(\mathcal{A})>0}\left\{
   \sum_{b_1,\cdots,b_M}{b \choose b_1,\cdots,b_M}\right.\\\notag
   &\quad\quad\quad\prod_{m=1}^M\left(C_{M}^m(-1)^{m+1}\right)^{b_m}\exp\bigg(-2\pi\lambda\frac{R_S}{R_E}\\\notag
   &\quad\quad\quad\left.\int_{r_1}^{R_{max}}\left(1-\prod_{m=1}^M\frac{1}{\left(1+\frac{m\eta\theta r_1^{\alpha}}{M r^{\alpha}}\right)^{Mb_m}}\right)r\,dr\bigg)\right\}\\\notag
   &\quad\quad\approx\mathbb{E}_{r_1|\Phi(\mathcal{A})>0}\left\{
   \sum_{b_1,\cdots,b_M}{b \choose b_1,\cdots,b_M}\right.\\\notag
   &\quad\quad\quad\left.\prod_{m=1}^M\left(C_{M}^m(-1)^{m+1}\right)^{b_m}\exp\left(-Q(r_1,\theta)\right)\right\},
\end{align}
where the last step is obtained by applying Gaussian-Chebyshev approximation to the integration of the exponent.

Finally, by taking expectation with respect to $r_1$ given $\Phi(\mathcal{A})>0$, and applying the Gaussian-Chebyshev approximation, the proof is complete.
\end{IEEEproof}

\begin{Remark}
Note that when $b=1$, $M_1(\theta)$ is the mean value of $P_s(\theta)$, which is the so called ``coverage probability''. The coverage probability can only reflect the average performance of all users in the considered scenario.
{\color{black}The variance of $P_s(\theta)$ can be obtained as $var(P_s(\theta))=M_2(\theta)-M_1^2(\theta)$. Note that, according to the ergodicity of $\Phi$, the larger the variance is, the larger the difference of the {\color{black}users'} obtained QoS becomes. Hence, the variance of $P_s(\theta)$ can be used to indicate user fairness of the considered scenario \cite{feng2019location}.}
\end{Remark}

The meta distribution of SIR is defined as:
\begin{align}
\bar{F}_{P_s}(\theta,x)\overset{\Delta}{=}\Pr\left(P_s(\theta)>x\right), x\in[0,1],
\end{align}
which is actually the CCDF of $P_s(\theta)$.
An exact integral expression for  $\bar{F}_{P_s}(\theta,x)$ can be formulated by applying
Gil-Pelaez inversion theorem \cite{haenggi2015meta}. However, it is very challenging to evaluate numerically and provide insights. Thus, in practice, $\bar{F}_{P_s}(\theta,x)$ is usually approximated as a beta distribution,
by matching the mean and
variance of the beta distribution with $M_1(\theta)$ and $M_2(\theta)$ \cite{haenggi2015meta}.

Note that, the PDF of a beta distributed
random variable $Z$, can be uniquely characterized by two parameters $\kappa$ and $\beta$, as follows:
\begin{align}
 f_{Z}(z)=\frac{z^{\kappa-1}(1-z)^{\beta-1}}{B(\kappa,\beta)},
\end{align}
where $B(\kappa,\beta)$ is the beta function, and
\begin{align}
 \mathbb{E}\{Z\}=\frac{\kappa}{\kappa+\beta}, \quad \mathbb{E}\{Z^2\}=\frac{\kappa(\kappa+1)}{\kappa+\beta(\kappa+\beta+1)}.
\end{align}

Let $\mathbb{E}\{Z\}=M_1(\theta)$ and $\mathbb{E}\{Z^2\}=M_2(\theta)$, it can be obtained that:
\begin{align}
 \kappa=\frac{M_1M_2-M_1^2}{M_1^2-M_2}, \text{ } \beta=\frac{(1-M_1)(M_2-M_1)}{M_1^2-M_2},
\end{align}
and the CCDF of $P_s(\theta)$ can be approximated as:
\begin{align}
 \bar{F}_{P_s}(\theta,x)\approx1-I_{x}(\kappa,\beta),
\end{align}
where $I_{x}(\kappa,\beta)$ is the regularized incomplete beta function, given by
$I_{x}(\kappa,\beta)=\frac{\int_{0}^{x}t^{\kappa-1}(1-t)^{\beta-1}\,dt}{B(\kappa,\beta)}$.
\section{Simulation results}
In this section, numerical results are presented to demonstrate performance of the considered LEO satellite communication system. The parameters are set as: $R_E=6371$ km, $\alpha=3.5$.
The simulations are obtained by taking average over $10000$ realizations of HPPP on the satellite sphere surface.

\begin{figure}[!t]
\vspace{-2em}
\setlength{\abovecaptionskip}{0em}   %
\setlength{\belowcaptionskip}{-1em}   %
  \centering
  \includegraphics[width=2.6in]{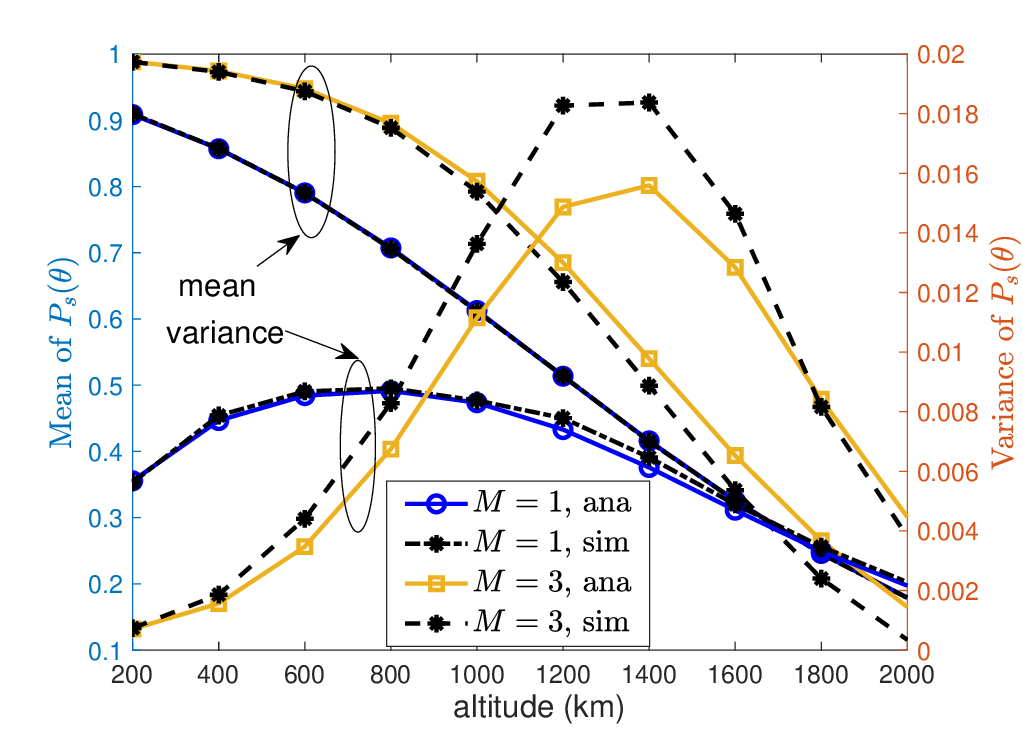}\\
  \caption{Mean and variance of $P_s(\theta)$. $\lambda=10^{-12}$, $\theta=0.1$. }
  \label{MV_1}
\end{figure}

\begin{figure}[!t]
\vspace{-0.5em}
\setlength{\abovecaptionskip}{0em}   %
\setlength{\belowcaptionskip}{-1em}   %
  \centering
  \includegraphics[width=2.6in]{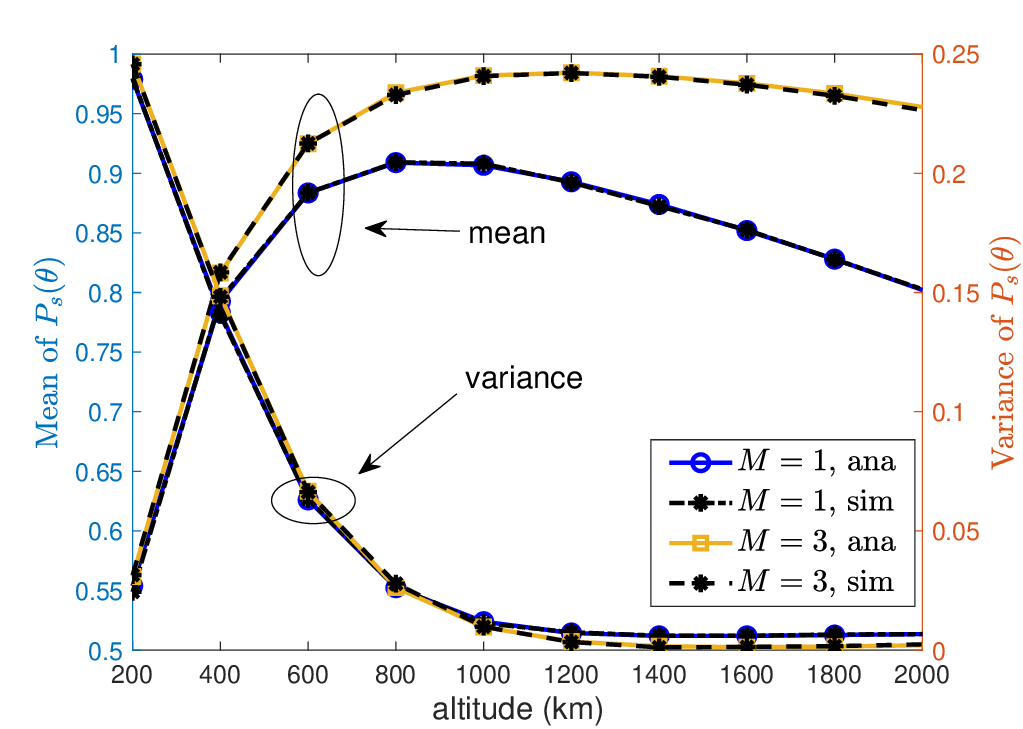}\\
  \caption{Mean and variance of $P_s(\theta)$. $\lambda=10^{-13}$, $\theta=0.1$. }
   \label{MV_2}
\end{figure}

Figs. \ref{MV_1} and \ref{MV_2} show the mean and variance of $P_s(\theta)$. As shown in Fig. \ref{MV_1}, when $\lambda=10^{-12}$, $M_1(\theta)$ decreases with the LEO altitude, while the variance first increases and then decreases. Thus, it can be easily concluded from Fig. \ref{MV_1} that: for $\lambda=10^{-12}$, it is beneficial to deploy  satellites at low attitudes, in terms of both coverage probability and user fairness. Differently, for a lower satellite density $\lambda=10^{-13}$, opposite trends are observed:
the mean value first increases and then slightly decreases with the LEO altitude, while the  variance decreases with the LEO altitude, and hence it is better to deploy satellites at middle and high altitudes.

\begin{figure}[!t]
\vspace{-0.5em}
\setlength{\abovecaptionskip}{-0em}   %
\setlength{\belowcaptionskip}{-1.8em}   %
  \centering
  \includegraphics[width=2.6in]{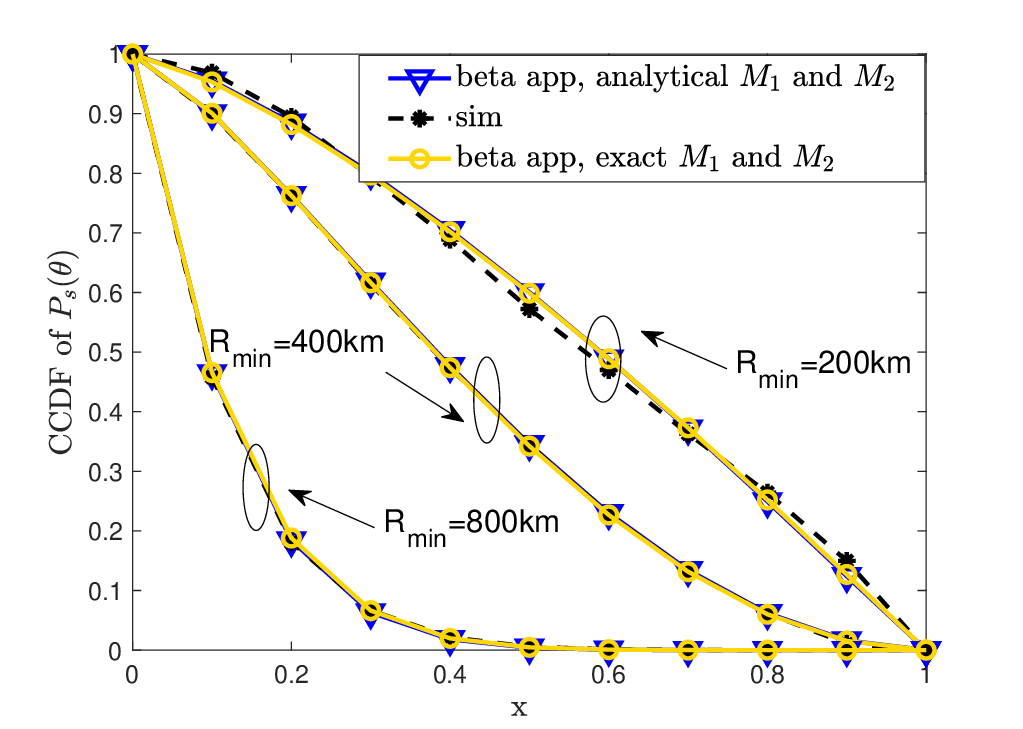}\\
  \caption{CCDF  of $P_s(\theta)$. $\lambda=10^{-12}$, $M=1$, $\theta=1$. }
  \label{meta_1}
\end{figure}
\begin{figure}[!t]
\vspace{-2em}
\setlength{\abovecaptionskip}{0em}   %
\setlength{\belowcaptionskip}{-1.8em}   %
  \centering
  \includegraphics[width=2.6in]{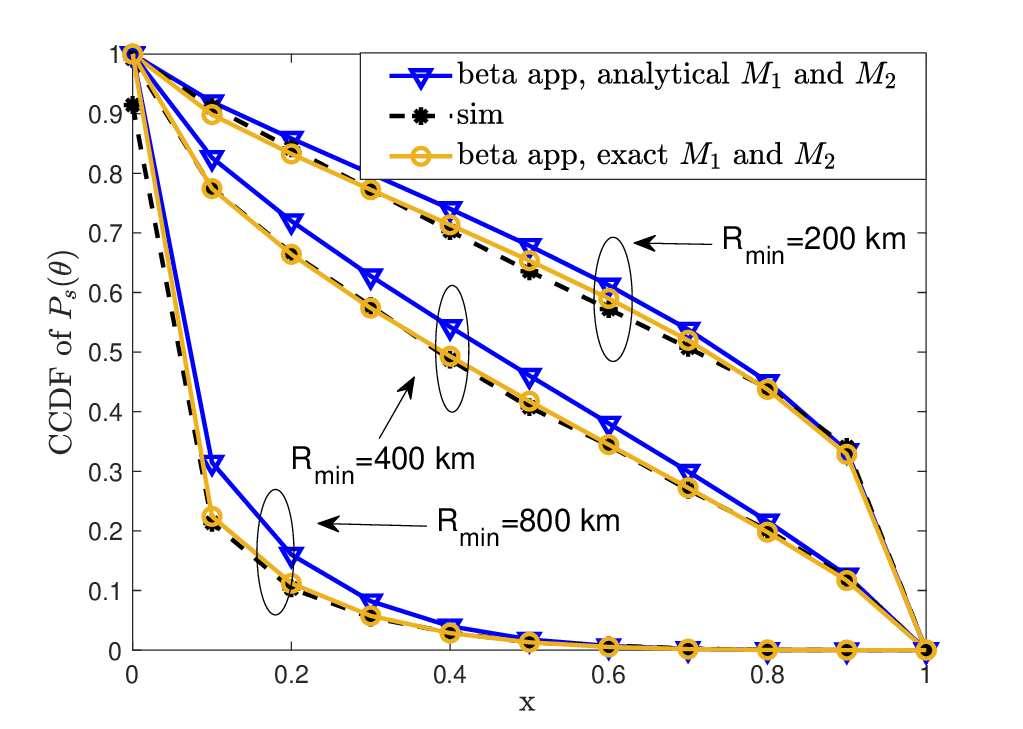}\\
  \caption{CCDF of $P_s(\theta)$. $\lambda=10^{-12}$, $M=3$, $\theta=1$. }
  \label{meta_2}
\end{figure}

Figs. \ref{meta_1} and \ref{meta_2} demonstrate the meta distribution of SIR.
Exact values of $M_1$ and $M_2$ are obtained via Monte Carlo simulations.
From the figures, it is shown that, analytical results perfectly match simulation results when $M=1$.
However, when $M=3$, analytical results cannot perfectly match simulation results, because $M_1(\theta)$
and $M_2(\theta)$ cannot be accurately evaluated as shown in Fig. \ref{MV_1}.
In addition, it can be observed from Figs. \ref{meta_1} and \ref{meta_2}, when $\lambda=10^{-12}$, it is better to deploy the satellite at LEO altitude $R_{min}=200$ km compared to higher altitudes $400$ km and
$800$ km, because for a given $0<x<1$, the CCDF of $P_s(\theta)$ at $R_{min}=200$ km is larger than those at
$R_{min}=400$ km and $R_{min}=800$ km. Note that the observation here is consistent with that shown in Fig.  \ref{MV_1}, as stated in the previous paragraph.
\vspace{-1.1em}
\section{Conclusion}
This paper has provided a fine grained stochastic geometry based performance
analysis on downlink LEO satellite communication networks.
HPPP has been applied to model the locations of satellites.
The moments and distribution of the conditional coverage probability given the point process have been investigated. It has been shown that, for a dense satellite constellation, it is better to deploy satellites at low altitude, by jointly considering users' average performance and user fairness, while for a sparse constellation, high altitude is favorable for satellite deployment.
\vspace{-0.8em}
\bibliographystyle{IEEEtran}
\bibliography{IEEEabrv,ref}
\end{document}